\documentclass[journal]{IEEEtran}
\usepackage{amsmath,amsfonts,amssymb}
\usepackage{graphicx}
\usepackage{setspace}
\usepackage{tocloft}
\usepackage{textgreek}
\usepackage[affil-it]{authblk} 
\usepackage{etoolbox}
\usepackage{lmodern}
\usepackage{authblk}
\usepackage{xcolor,soul,framed} 
\usepackage{lipsum}
\newcommand\blfootnote[1]{%
  \begingroup
  \renewcommand\thefootnote{}\footnote{#1}%
  \addtocounter{footnote}{-1}%
  \endgroup
}

\usepackage{amssymb}

\usepackage{subcaption}
\usepackage[symbol]{footmisc}
\renewcommand{\thefootnote}{\fnsymbol{footnote}}

\newcounter{daggerfootnote}

\colorlet{shadecolor}{yellow}

\usepackage{array}
\usepackage{mdwmath}
\usepackage{mdwtab}
\usepackage{eqparbox}
\usepackage{url}

\begin{document}
\bstctlcite{IEEEexample:BSTcontrol}
\title{Nanobeacon: A time calibration device for the KM3NeT neutrino telescope}
    
\author[a]{S.~Aiello}
\author[ba,b]{A.~Albert}
\author[c]{M.~Alshamsi}
\author[d]{S. Alves Garre}
\author[e]{Z.~Aly}
\author[f,g]{A. Ambrosone}
\author[h]{F.~Ameli}
\author[i]{M.~Andre}
\author[j]{G.~Androulakis$^\dagger$\footnote[2]{† Deceased}}
\author[k]{M.~Anghinolfi}
% ...

\affil[a]{INFN, Sezione di Catania, Via Santa Sofia 64, Catania, 95123 Italy}
\affil[b]{Universit{\'e}~de~Strasbourg,~CNRS,~IPHC~UMR~7178,~F-67000~Strasbourg,~France}
\affil[c]{Universit{\'e} de Paris, CNRS, Astroparticule et Cosmologie, F-75013 Paris, France}
\affil[d]{IFIC - Instituto de F{\'\i}sica Corpuscular (CSIC - Universitat de Val{\`e}ncia), c/Catedr{\'a}tico Jos{\'e} Beltr{\'a}n, 2, 46980 Paterna, Valencia, Spain}
\affil[e]{Aix~Marseille~Univ,~CNRS/IN2P3,~CPPM,~Marseille,~France}
\affil[f]{INFN, Sezione di Napoli, Complesso Universitario di Monte S. Angelo, Via Cintia ed. G, Napoli, 80126 Italy}
\affil[g]{Universit{\`a} di Napoli ``Federico II'', Dip. Scienze Fisiche ``E. Pancini'', Complesso Universitario di Monte S. Angelo, Via Cintia ed. G, Napoli, 80126 Italy}
\affil[h]{INFN, Sezione di Roma, Piazzale Aldo Moro 2, Roma, 00185 Italy}
\affil[i]{Universitat Polit{\`e}cnica de Catalunya, Laboratori d'Aplicacions Bioac{\'u}stiques, Centre Tecnol{\`o}gic de Vilanova i la Geltr{\'u}, Avda. Rambla Exposici{\'o}, s/n, Vilanova i la Geltr{\'u}, 08800 Spain}
\affil[j]{NCSR Demokritos, Institute of Nuclear and Particle Physics, Ag. Paraskevi Attikis, Athens, 15310 Greece}
\affil[k]{INFN, Sezione di Genova, Via Dodecaneso 33, Genova, 16146 Italy}
% ...
\affil[ba]{Universit{\'e} de Haute Alsace, rue des Fr{\`e}res Lumi{\`e}re, 68093 Mulhouse Cedex, France}
% ...


\markboth{Arxiv
}{Real \MakeLowercase{\textit{et al.}}: Nanobeacon: A time calibration device for the KM3NeT neutrino telescope}

\begin{titlepage}
\maketitle
\end{titlepage}

\begin{abstract}
The KM3NeT Collaboration is currently constructing a multi-site high-energy neutrino telescope in the Mediterranean Sea consisting of matrices of pressure-resistant glass spheres, each holding a set of 31 small-area photomultipliers. The main goals of the telescope are the observation of neutrino sources in the Universe and the measurement of the neutrino oscillation parameters with atmospheric neutrinos. Both extraterrestrial and atmospheric neutrinos are detected through the Cherenkov light induced in seawater by charged particles produced in neutrino interactions in the surrounding medium. A relative time synchronization between photomultipliers of the order of 1 ns is needed to guarantee the required angular resolution of the detector. Due to the large detector volumes to be instrumented by KM3NeT, a cost reduction of the different systems is a priority. To this end, the inexpensive Nanobeacon  has been designed and developed by the KM3NeT Collaboration to be used for detector time-calibration studies. At present, more than 600 Nanobeacons have been already produced. The characterization of the optical pulse and the wavelength emission profile of the devices are critical for the time calibration. In this paper, the main features of the Nanobeacon design, production and operation, together with the main properties of the light pulse generated are described.   \end{abstract}


%
\IEEEpeerreviewmaketitle


\section{Introduction}  \label {sec:intro}

\IEEEPARstart{T}{he} KM3NeT Collaboration~\cite{km3net_letter} is building a network of underwater neutrino telescopes at two deep locations in the Mediterranean Sea. Although the same technology is used for the two detectors their configuration is different, reflecting the difference in scientific goals. Astroparticle Research with Cosmics in the Abyss (ARCA~\cite{ARCA}) has been designed for the detection of neutrinos of astrophysical origin with energies from $\sim$ 100 GeV to PeV, and is located 100 km off the southern tip of Sicily (Italy) at a depth of 3500 m. Oscillation Research with Cosmics in the Abyss (ORCA~\cite{ORCA}), which mainly aims at studying the fundamental properties of neutrinos, is located 40 km south of the coast of Toulon (France) at a depth of about 2450 m.  The detection principle is based on the collection of Cherenkov photons produced along the path of relativistic charged particles emerging from neutrino interactions inside or in the vicinity of the detector.

\begin{figure}
\begin{center}
\includegraphics[width=3.5in]{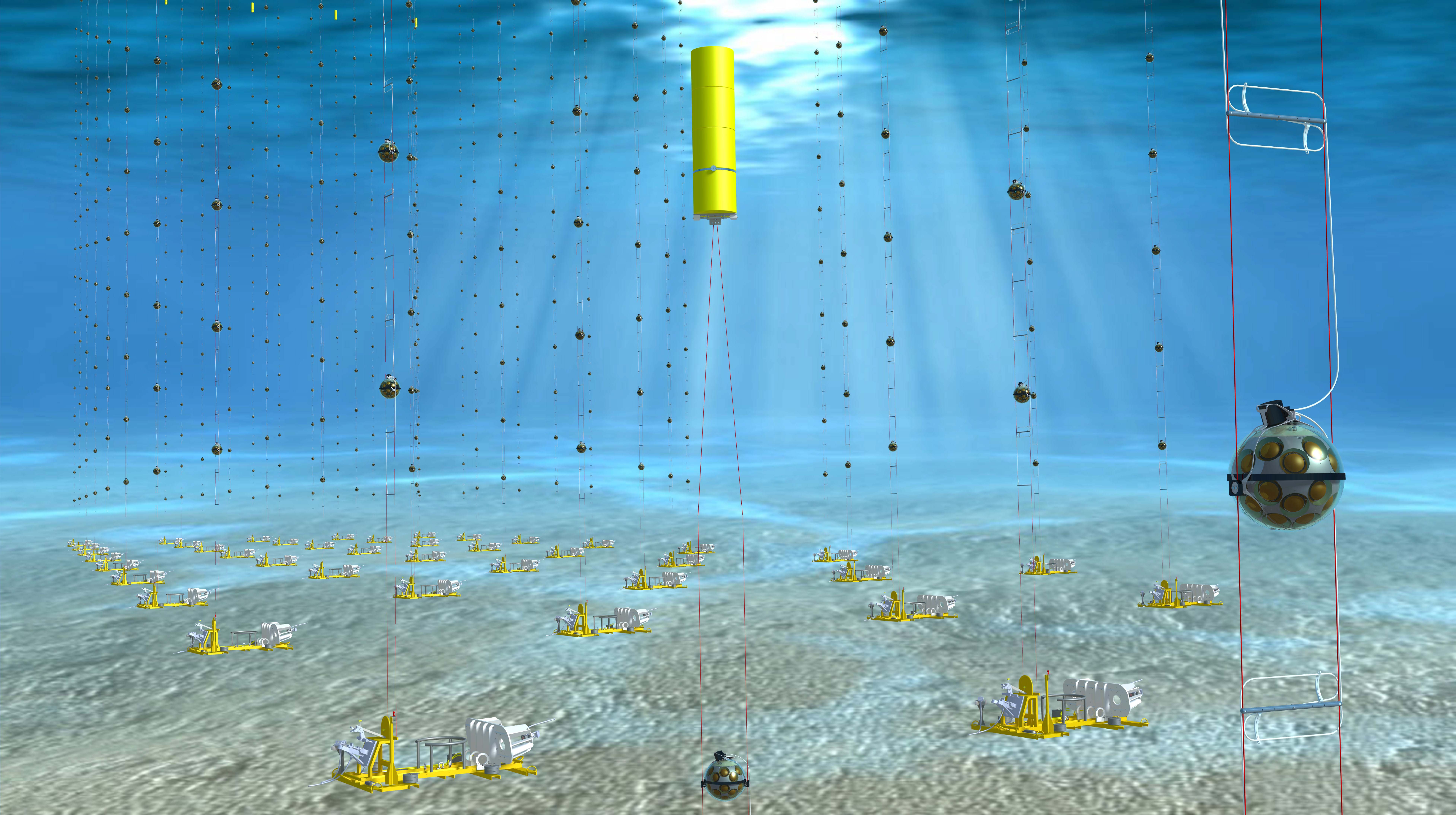}
\caption{Artistic view of KM3NeT. The illustration is not scaled: sunlight does not reach the depths at which the KM3NeT detector is deployed. The total volume of the detector, once completed, will be larger than one km\textsuperscript{3}.}
 
\label{fig:artist}
\end{center}
\end{figure}

\blfootnote{E-mail addresses: real@ific.uv.es, dacaldia@ific.uv.es, sagreus@ific.uv.es, km3net-pc@km3net.de}
The telescopes are designed as 3-dimensional matrices of light detectors, spanning a large volume, as shown in Fig.~\ref{fig:artist}. The active component is the Digital Optical Module (DOM)~\cite{DOM1,DOM2,DOM3}, a pressure-resistant glass sphere which houses 31 photomultipliers (PMTs). The arrival time of the Cherenkov photons, together with the position of the DOMs allow the reconstruction of the trajectory of the primary neutrino. The resolution of the reconstructed neutrino trajectory in the detector depends on the accurate measurement of the arrival time of the light on the optical sensors, with a precision better than one nanosecond, and of their position in space, better than 20 cm. For that reason, time and position calibration of the telescope is a critical requirement. The main data acquisition electronic board of the DOM~\cite{elec1,elec2} is the Central Logic Board (CLB)~\cite{clb, clb1}, which performs the readout of the 31 PMT channels. A power board~\cite{pb} mounted on top of the CLB provides power. The DOMs are distributed along lines, called Detection Units (DUs), each hosting 18 DOMs. In ORCA, the vertical spacing between DOMs is around 9~m, while in ARCA it is 36~m. The DUs are anchored on the seafloor, with a spacing about 20 m in ORCA, and 90 m in ARCA and are organized in Building Blocks (BB), each one composed of 115 DUs. The first fourteen DUs of KM3NeT, six for ORCA and eight for ARCA, have already been deployed and are currently taking data~\cite{dependence}.

\subsection{The Nanobeacon system} \label {sec:instru}
In addition to the information encoded in the clock distribution time using a customized version~\cite{km3netcalib} of the White Rabbit protocol~\cite{wr1,wr2}, there are several methods of \textit{in-situ} time calibration. In addition to down-going muons and photons from $^{40}$K radioactive decays in the seawater, KM3NeT has an optical beacon system based on pulse generation devices. Two different sources of short-pulsed light~\cite{timecalib} are used to cross-check the time calibration of the telescope by illuminating sets of the optical modules in a controlled scenario: the laser beacon ~\cite{laser1}, and the Nanobeacon~\cite{Nano}, the latter being the subject of the present work. 

\begin{figure}[tbp] 
\begin{center}

\begin{subfigure}[p]{.6\linewidth}
\includegraphics[trim={3cm 1cm 0cm 3cm},width=1\textwidth]{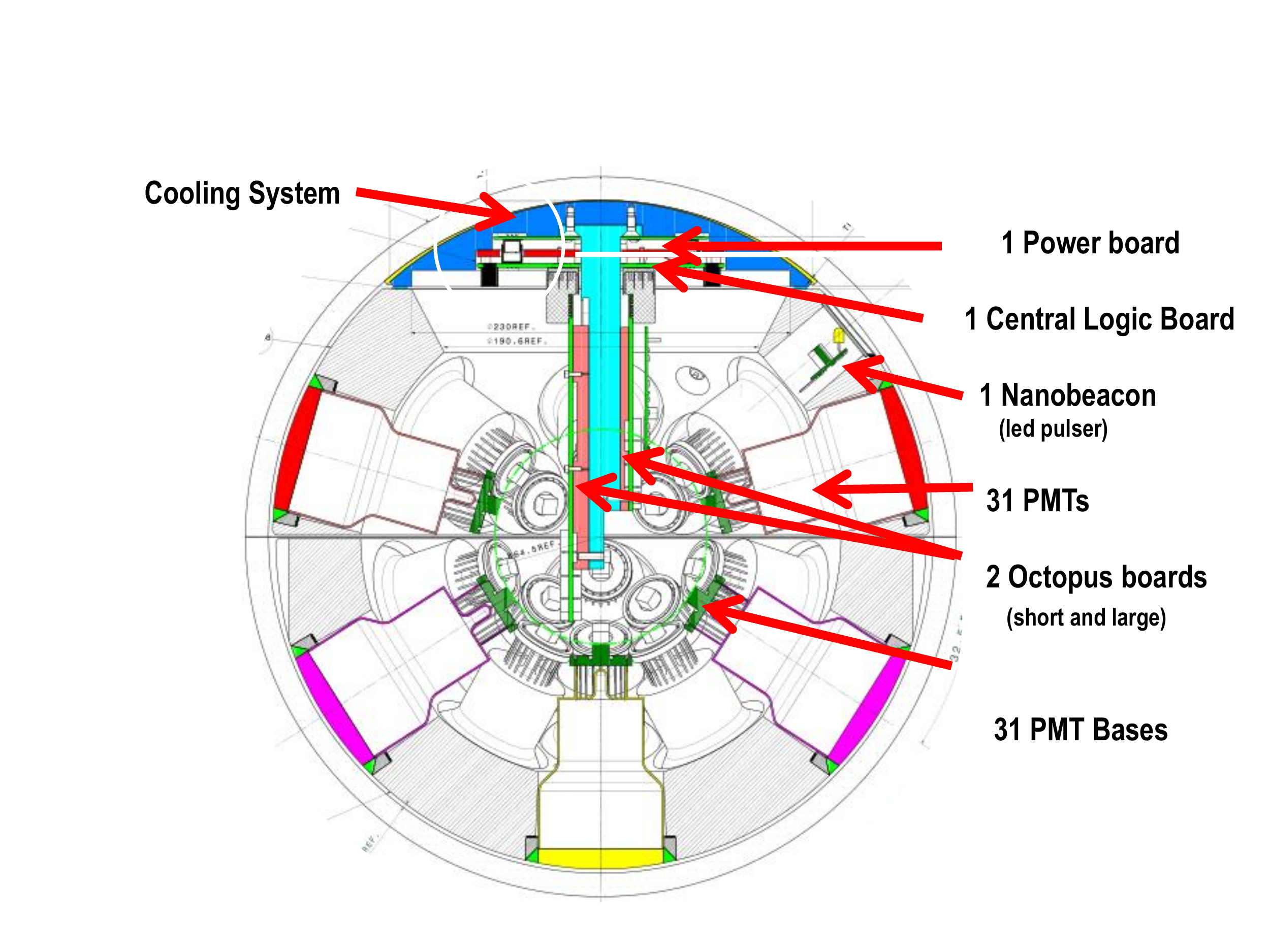}  
\caption{}
 \label{fig:overview}
\end{subfigure}
\begin{subfigure}[p]{.35\linewidth}
\includegraphics[trim={6cm 0cm 6cm 0cm},width=1\textwidth]{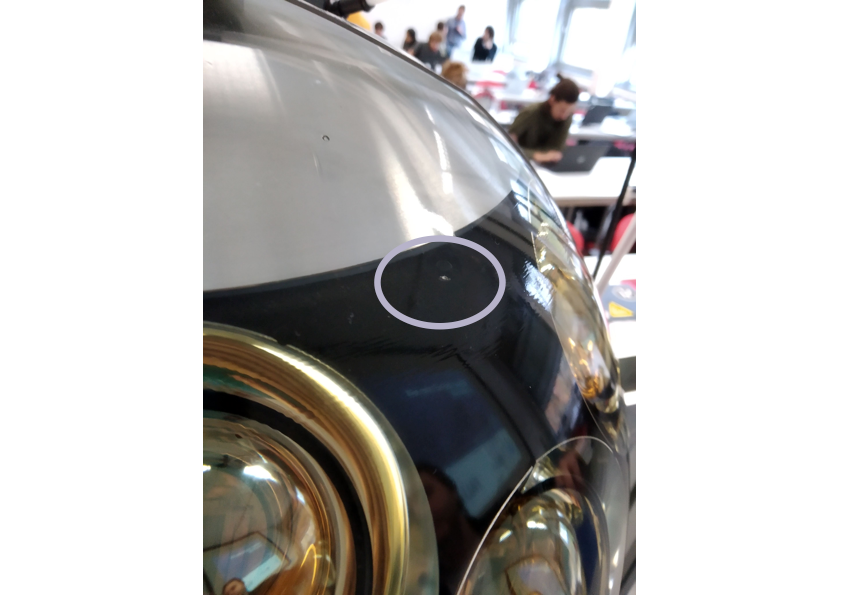}  
\caption{}
 \label{fig:domzoom}
\end{subfigure}

\caption{(a) 2D sketch of the DOM, showing the position of the Nanobeacon LED at 45$^{\circ}$ from the vertical axis. The main components of the DOMs are shown: the PMTs, the cooling system and the main electronic boards.  (b) A picture of a small part of the top hemisphere of the KM3NeT optical module with the aluminum cooling cap visible. The DOM sphere has a diameter of 17'' (43 cm). The position of the Nanobeacon is highlighted by the circle.}

\label{fig:dom}
\end{center}
\end{figure}

Each DOM houses a Nanobeacon (Fig.~\ref{fig:dom}). The Nanobeacon can produce a short duration pulse approximately 5~ns wide (FWHM), with around 3~ns rise time (from 10\% to 90\% of full amplitude) and with enough intensity to illuminate the DOMs located on the same DU above the emitting DOM. With a good understanding of the optical pulse shape and thanks to the control of the time of emission, the Nanobeacon pulses can be used to monitor the relative time offsets between DOMs of the same DU.


The intensity of the pulse and the angular aperture (with a solid angle with 50\% of light emission) of the LED emission have also been proven relevant for the calibration between DUs, as a proper selection of these values allows the illumination of the neighboring DUs with Nanobeacon flashes. The position of the Nanobeacon 45$^{\circ}$  off the axis from the top DOM is chosen to reduce the effect of sedimentation and biofouling on the glass sphere of the DOM.

The electronics and the control firmware of the Nanobeacon are presented in Sections~\ref{sec:elec} and ~\ref{sec:firmware}, respectively, while studies of the LED model are addressed in Section~\ref{sec:leds}. The mass production of Nanobeacons for KM3NeT is described in Section~\ref{sec:manufacturing} , together with the test setup and the results of the intensity tests performed just after production. The proof of concept is illustrated in Section ~\ref{sec:proof}, while the characterization of the optical pulse in terms of rise time and width are presented in Section~\ref{sec:optical}. Moreover, the results of the pulse characterization in terms of wavelength are discussed in this Section. Finally, in Section~\ref{sec:conclu}, a summary is presented together with future plans.

\section{Nanobeacon electronics}\label {sec:elec}
The Nanobeacon has its origin in the LED beacon system used for the time calibration of the Astronomy with a Neutrino Telescope and Abyss environmental RESearch (ANTARES) experiment~\cite{ABeacon}. ANTARES, which is still operational after more than 13 years in the Mediterranean Sea close to the ORCA site, is the predecessor of KM3NeT and is based on the same detection principle but with smaller instrumented volume.
The ANTARES LED beacon consists of a glass and titanium container housing 36 LEDs with their corresponding electronic pulsers. The LED beacon  has been considerably simplified in KM3NeT by integrating the LED inside the DOM. This avoids the need for a mechanical external container and reduces significantly the cost. The Nanobeacon is composed of two different elements: the pulser, and the DC/DC converter. The pulser comprises the LED and the electronics board that generates the trigger for the optical pulse. The DC/DC converter is integrated in the power board inside the DOM, which generates the power supply to the Nanobeacon pulser. The power generated in the power board is fed to the Nanobeacon via the CLB. The control of the rail and the generation of the electrical trigger are both carried out at the CLB.

\subsection{Nanobeacon pulser} 

\begin{figure}[tbp] 
\begin{center}

\begin{subfigure}[p]{1\linewidth}
\includegraphics[angle=-0,origin=c,trim={0cm 0cm 0cm 0},clip,width=1\textwidth]{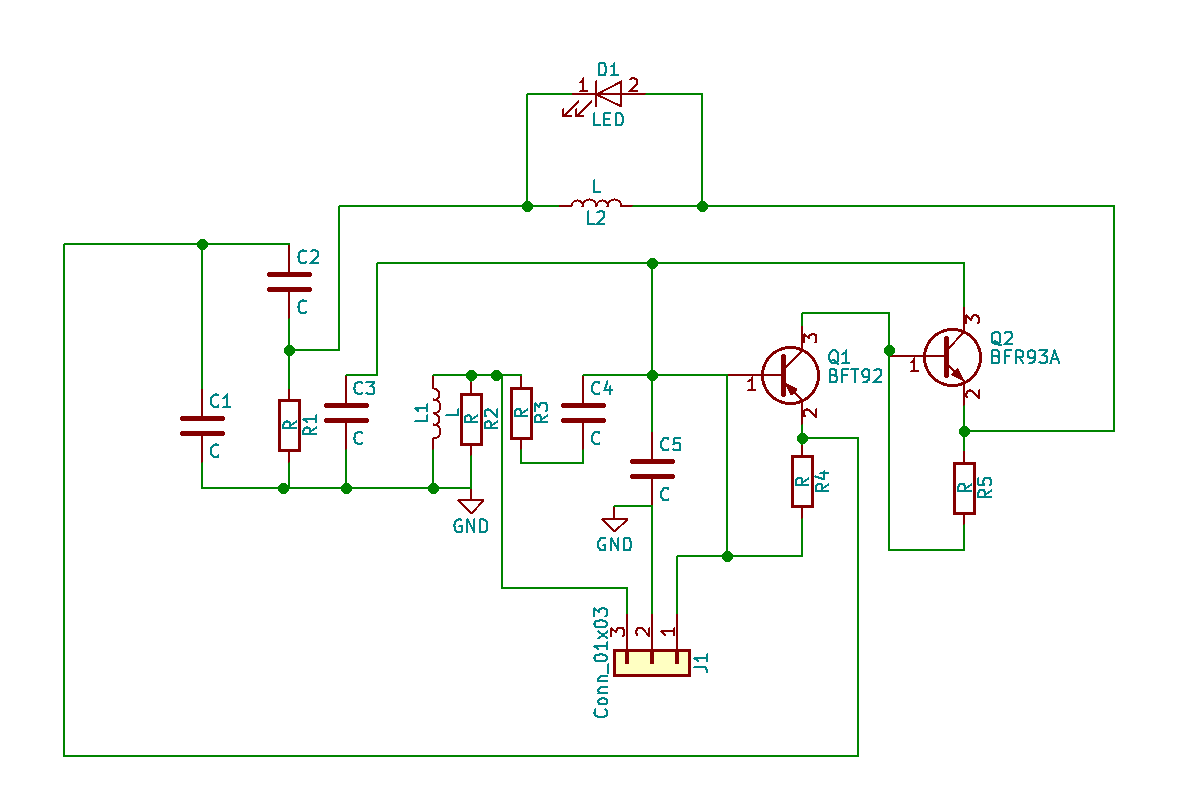}  
\caption{}
 \label{fig:nsc}
\end{subfigure}
\begin{subfigure}[p]{1\linewidth}
\includegraphics[angle=0,origin=c,trim={0cm 0cm 0cm 0},clip,width=1\textwidth]{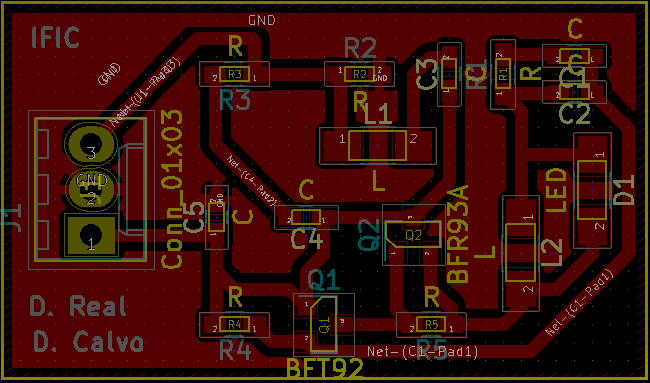}
\caption{}
 \label{fig:pulser}
\end{subfigure}

\caption{(a) Electronics schematics of the Nanobeacon pulser.  (b) Layout of the Nanobeacon pulser. The dimensions of the Nanobeacon pulser are 35 mm $\times$ 20 mm.}
\label{fig:elec_nano}
\end{center}
\end{figure}

The Nanobeacon pulser is based on the Kapustinsky circuit~\cite{Kapu}, which can generate short pulses with simple electronics, modified for KM3NeT. The schematics of the pulser are shown in Fig.~\ref{fig:nsc} while the layout is presented in Fig.~\ref{fig:pulser}. The trigger of the pulser is generated by an FPGA on the CLB, which provides a 1.5 V negative square pulse. The pulse generated is added to the DC voltage  provided by the power supply rail.  The DC voltage loads a capacitor (C2) that is discharged to the LED during the transition of two transistors working in opposition. The switching between transistors is triggered by the signal provided by the FPGA, provoking the discharge of the capacitor to the LED and the emission of a narrow optical pulse. The trigger generated by the FPGA is provided to the Nanobeacon pulser by a three-wire twisted cable to reduce the electromagnetic induced noise. This connection also provides ground and power, which comes from the power board. 

\subsection{Nanobeacon power supply} 


The Nanobeacon power supply is provided by one of the rails of the power board, which converts the 12 V input into the voltage needed to operate the Nanobeacon.  The power rail consists of a DC/DC converter working in a buck-boost configuration, providing a voltage configurable between 4.5 and 30 V, via a DAC controllable by an I\textsuperscript{2}C. When the output voltage of the DAC changes, the voltage of the Nanobeacon power supply is modified accordingly. The provided voltage determines the amount of current supplied to the LED, and therefore, the intensity of the generated optical pulse. The power supply includes two circuits for measuring the voltage and the current. The values provided by these circuits are acquired with an ADC that is read by the FPGA embedded software. As will be shown in Section~\ref{sec:optical}, the voltage is correlated with the pulse width: the lower the voltage supplied, the narrower the optical pulse. The typical power consumption of the Nanobeacon does not exceed  0.1 W.

\section{Firmware}\label {sec:firmware}

\begin{figure}[tbp] 

\begin{center}
\includegraphics[angle=0,origin=c,trim={3cm 6cm 8cm 3cm},clip,width=.5\textwidth]{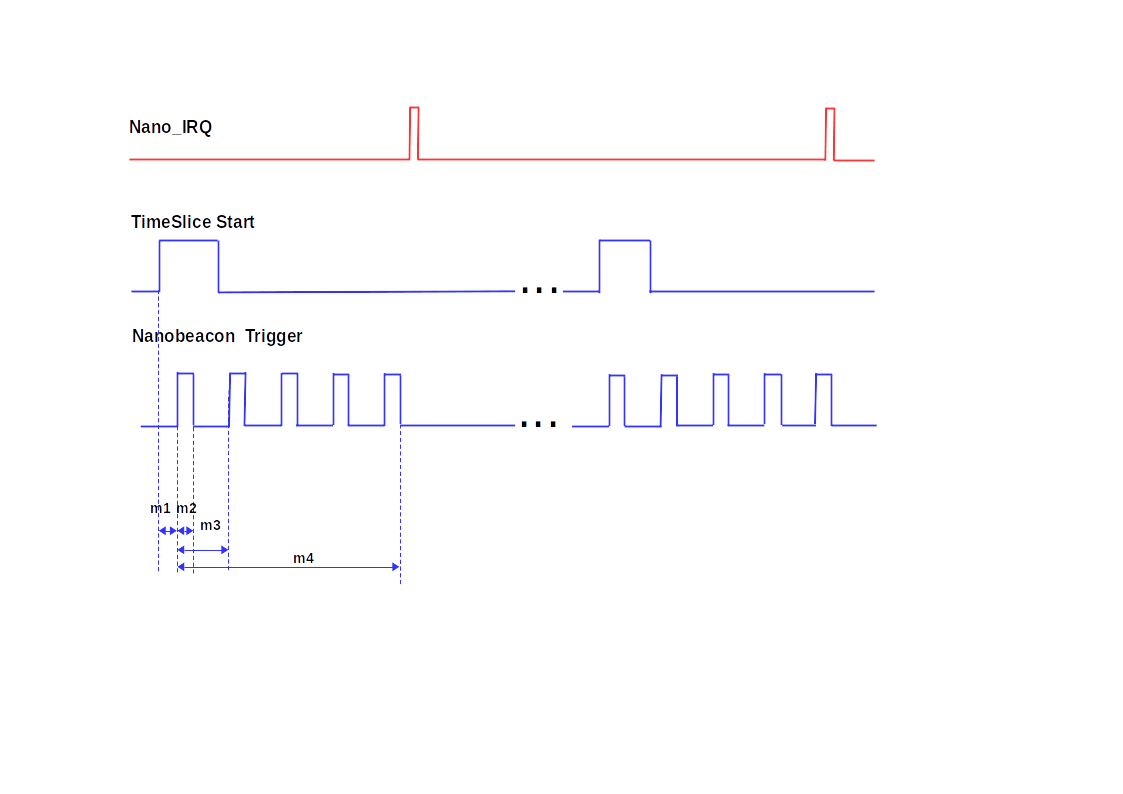}
\caption{Scheme of the Nanobeacon firmware IP core working operation. The Nanobeacon trigger is defined by the delay after the start of the timeslice (m1, and coded in the second register), the number of pulses per timeslice (m4, and coded in the fifth register), the duty cycle of the pulse (m2, and coded in the third register) and its frequency (m3, and coded as period in the fourth register). The interrupt (Nano\_IRQ) generated by the firmware IP core when the number of pulses per timeslice has ended is also shown.}
\label{fig:nfi}
\end{center}
\end{figure}

An Intellectual Property (IP) block, implemented in the CLB FPGA, generates the trigger signal for the Nanobeacon. This FPGA IP core implements a Wishbone slave~\cite{wish}, where 5 registers of 32 bits are accessible via the embedded software to configure the Nanobeacon trigger signal  (Table~\ref{tab:tbnbr}). In particular, the first register  allows for enabling and disabling the trigger signal.  For this, one of the bits of the first register, called \textit{Enable control}, is used. Another bit, called \textit{Enable power}, controls the power supply of the Nanobeacon. There are, therefore, two different  methods for disabling the Nanobeacon.  In KM3NeT, the data acquisition is organized in \textit{timeslices}, which have a typical duration of 100 ms. A second register (m1) allows for defining the delay of the first flash after the start of the timeslices. A third register (m2) defines the Nanobeacon pulse length. The flash period is configured in a fourth register (m3) and the number of triggers per timeslice is defined in the fifth one (m4). In all cases, the time is coded in ticks of 16 ns. Table~\ref{tab:tbnbr}  describes the different control registers that configure the IP core to trigger the Nanobeacon pulse. The  operation of the Nanobeacon firmware is illustrated in Fig.~\ref{fig:nfi}. The configuration of the last 4 registers defines the characteristics of the flash rate, while the control of the pulse is implemented in the first register. Before starting the LED flashing, an I\textsuperscript{2}C command is sent via embedded software to the Nanobeacon rail to set the voltage. In addition to the configuration, the IP core sends an interrupt to the embedded software after the end of each flashing pattern in a timeslice has ended, giving feedback of the operation performed. 


\begin{table*}[tbp]
\caption{Control registers of the Nanobeacon IP core. The first register includes the two control enabling/disabling bits, while the remaining registers parameterize the pattern of the Nanobeacon trigger.}
\label{tab:tbnbr}
\smallskip
\centering
\begin{tabular}{|p{3cm}|p{3cm}|p{3cm}|}

  \hline 
   \multicolumn {1}{|c|}{ Reg} & \multicolumn {1}{|c|}{ Name}& \multicolumn {1}{|c|}{ Remark}\\
\hline 
  \hline 
   \multicolumn {1}{|c|}{0x00}& \multicolumn {1}{|c|}{Enable register }& \multicolumn {1}{|c|}{Bit 0:Enable control - Bit 1:Enable power - Bits(2-31):Not used}\\

 \hline 
  \multicolumn {1}{|c|}{ 0x01}& \multicolumn {1}{|c|}{Delay after TimeSlice}& \multicolumn {1}{|c|}{ In 16 ns ticks (default 0 ns)}\\

 \hline 
  \multicolumn {1}{|c|}{ 0x02}& \multicolumn {1}{|c|}{Nanobeacon Pulse Width}& \multicolumn {1}{|c|}{ In 16 ns ticks (default 64 ns)}\\

 \hline 
  \multicolumn {1}{|c|}{ 0x03}& \multicolumn {1}{|c|}{Nanobeacon Pulse Period}& \multicolumn {1}{|c|}{ In 16 ns ticks (default 50 \textmu s)}\\

 \hline 
  \multicolumn {1}{|c|}{ 0x04}& \multicolumn {1}{|c|}{Number of pulses}& \multicolumn {1}{|c|}{ Number of pulses (default 100 pulses)}\\

 \hline
\end{tabular}
\end{table*}

\section{Comparison of the LED model candidates} \label {sec:leds}

\begin{figure}[tbp] 
\begin{center}
\includegraphics[angle=0,origin=c,trim={3cm 0cm 0cm 0cm},clip,width=0.54\textwidth]{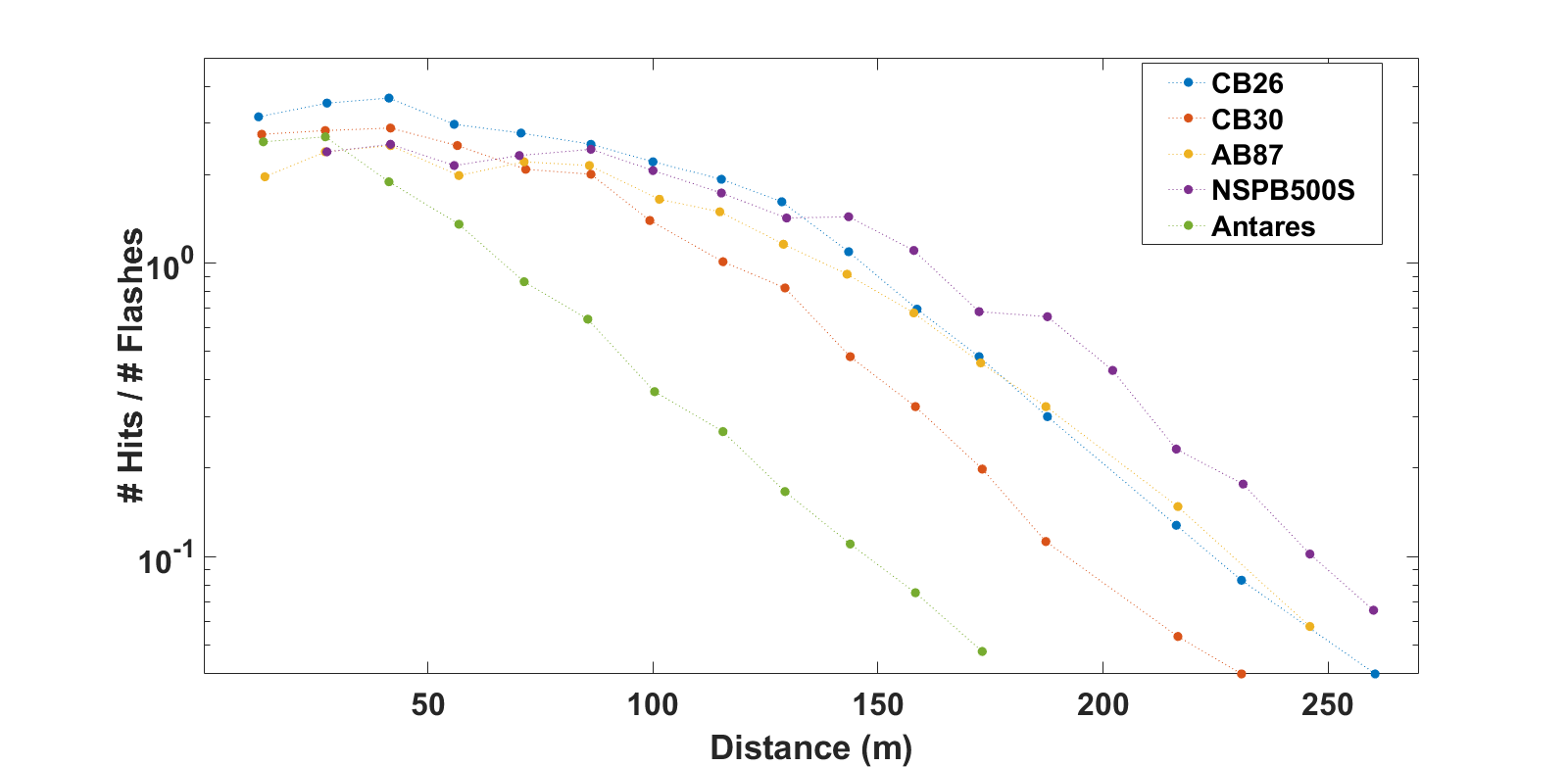}
\caption{Results of the \textit{in-situ} tests in one of the ANTARES lines of the preselected KM3NeT LEDs. The average number of hits recorded in each storey per flash is shown as a function of the distance from the flashing LED. Each point represents one storey, the stories are spaced 15 m apart. The final selected LED model for KM3NeT (HLMP-CB1A-XY0DD) is the evolution of CB26 and CB30.} 

\label{fig:antares_led}
\end{center}
\end{figure}

Several LED models (AVAGO HLMP-CB26, AVAGO HLMP-CB30, AGILENT HLMP-AB87, and Nichia NSPB500S) were tested in the laboratory and \textit{in-situ} in one of the ANTARES lines. The angular aperture of the models analyzed  ranges from 10º to 30º. The light intensity is one order of magnitude higher than the intensity of the LED model used in ANTARES, where the top part of the LEDs used was removed. The results of the tests performed using the ANTARES line are shown in Fig.~\ref{fig:antares_led}. Each point represents the average number of hits per flash detected on a storey, which in the case of the ANTARES test were spaced by approximately 15 m. The results are presented for all LED models under study and compared to the original ANTARES LED beacon. The target storey situated at 150 m from the emitting beacon detects approximately 0.1 hit per flash emitted by the ANTARES LED beacon, while other LED models considered for KM3NeT record values ranging from 0.4 to 1.05 hits per flash. The angular distribution of the analyzed LED models is wide enough to perform intraline calibration, i.e., to illuminate the DOMs of the same DU. As already mentioned, the first KM3NeT tests have shown that it is also possible to illuminate the DOMs of the nearest DUs, even in the case of ARCA where the lines are spaced 90 m away. The final LED model chosen for KM3NeT is the HLMP-CB1A-XY0DD by Broadcom. This model is the evolution of the two LEDs from Avago (CB26 and CB30) tested on the ANTARES line, which were obsolete and not available at the moment of the final decision. The main characteristics of this LED are summarized in Table~\ref{tab:tled}.  Its light intensity is more than twice the intensity of former models.

\begin{figure}[tbp] 
\begin{center}
\includegraphics[angle=0,origin=c,trim={6cm 1cm 6cm 1cm},clip,width=.49\textwidth]{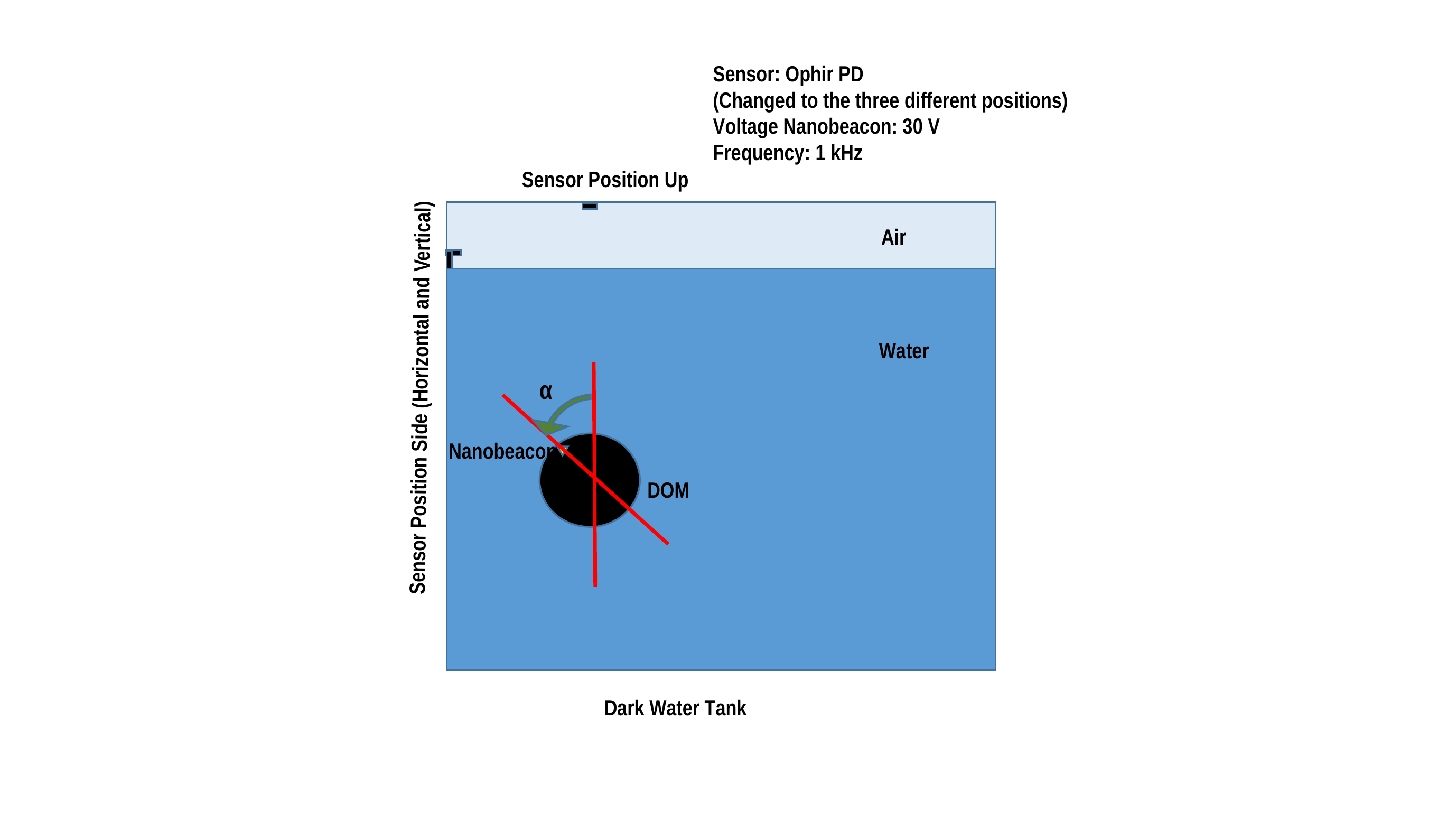}
\caption{Light-tight water tank setup. A DOM is submerged in the tank at a depth of approximately 2 m. The DOM can rotate, as shown in the figure. Three light sensors are located on the upper part of the container, collecting the light emitted by the Nanobeacon being flashed in the DOM.}

\label{fig:gamelle}
\end{center}
\end{figure}

\begin{figure}[tbp] 

\begin{center}
\includegraphics[angle=0,origin=c,trim={1.5cm 9.5cm 0cm 10cm},clip,width=.54\textwidth]{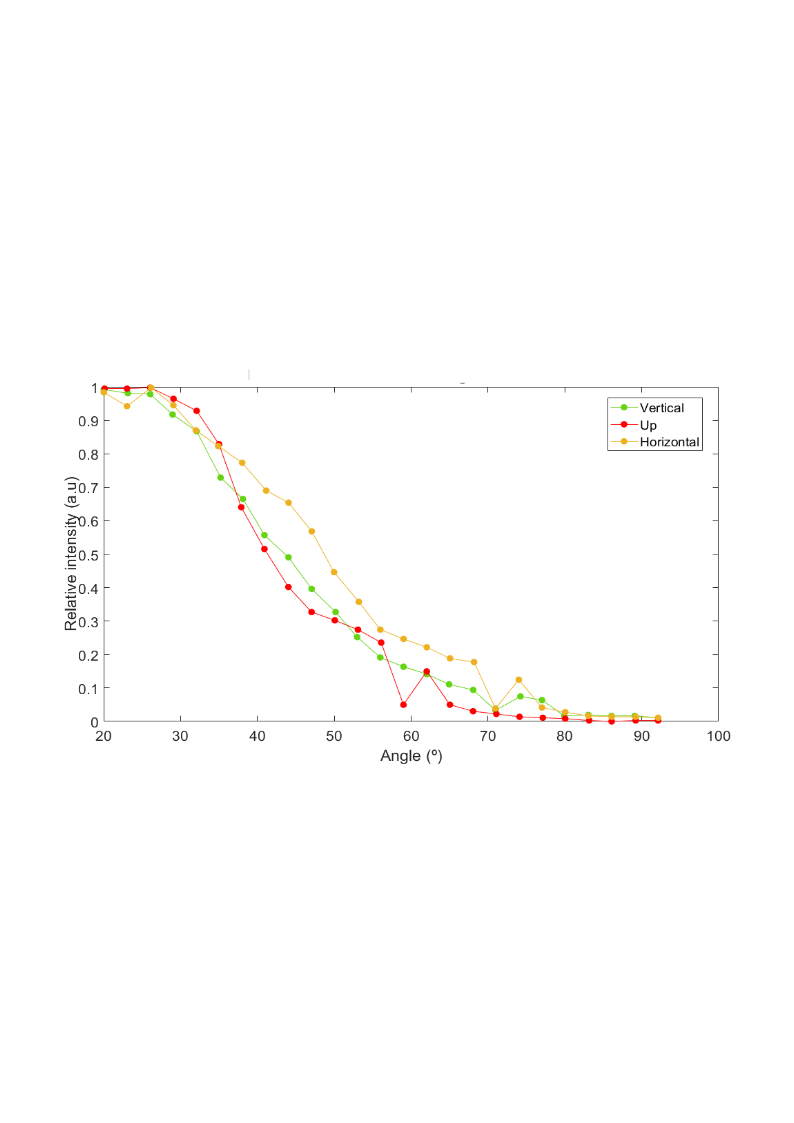}
\caption{Light  intensity as a function of the polar angle  measured in the water tank setup with three light sensors (see details in the text). Data have been normalized to the maximum light intensity.}

\label{fig:nanomeasurements}
\end{center}
\end{figure}

According to the manufacturer, the effective aperture of the chosen LED model is 15º in air. However, the angular aperture is modified by the light transitions through materials with different refraction indices, such as the DOM glass sphere and the seawater. In order to measure the real angular aperture, a light-tight water tank available at the Laboratoire AstroParticule \& Cosmologie, Paris, France (APC) was used. A DOM equipped with an embedded Nanobeacon was immersed in the water tank as shown in Fig.~\ref{fig:gamelle}. The DOM was mounted in a mechanical structure allowing its rotation. During the test, the water tank was completely closed and darkened. Three light sensors were installed on the inner walls of the water tank: two were located on one corner of the water tank, about 10 cm below the tank cover and right above the water surface. One of them was facing down while the other one was facing sideways. The third sensor was installed at the top cover of the tank, over the vertical axis of the DOM and 10 cm over the water surface. The three sensors were recording the measured light intensity while the DOM with the flashing LED was being rotated. Fig.~\ref{fig:nanomeasurements} shows the data read out for different angle positions. The data have been normalized to the maximum value. The angular aperture of the selected LED model in water exceeds 40º, much larger than the nominal value in air.

\begin{table}[tbp]
\renewcommand{\arraystretch}{1.3}
\caption{Main characteristics of the HLMP-CB1A-XY0DD LED model.}
\label{tab:tled}
\centering
\begin{tabular}{|c||c|}

  \hline 
 \multicolumn{1}{|c|}{ Attribute} & \multicolumn {1}{|c|}{ Value}\\
\hline 
\hline 
  \hline 

\multicolumn{1}{|c|}{ Color} & \multicolumn {1}{|c|}{ blue}\\
 \hline 
 \multicolumn{1}{|c|}{ Wavelength (nm) } & \multicolumn {1}{|c|}{ 470}\\

 \hline 
\multicolumn{1}{|c|}{ Aperture (deg) } & \multicolumn {1}{|c|}{ 15}\\

      \hline 
\multicolumn{1}{|c|}{ Luminous intensity (mcd) } & \multicolumn {1}{|c|}{ 12000}\\                     
 \hline 
\multicolumn{1}{|c|}{ Package } & \multicolumn {1}{|c|}{ T-1 3/4 (5 mm)}\\
 \hline
\end{tabular}
\end{table}

\section{Manufacturing and tests} \label {sec:manufacturing}

\begin{figure}[tbp] 
\begin{center}
\includegraphics[trim={0cm -2cm 0cm 0},clip,width=.499\textwidth]{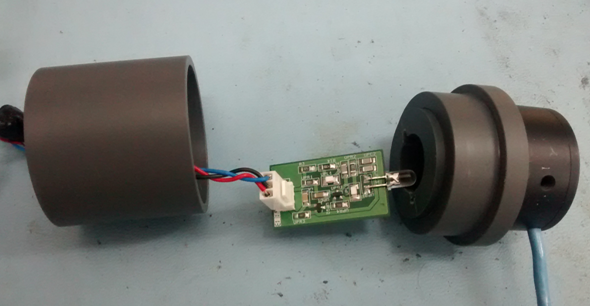}
\caption{Setup used for the production tests. Note the Nanobeacon between the two PVC pieces and the head of the intensity meter coupled with the PVC piece on the right.}

\label{fig:tna}
\end{center}
\end{figure}

 The first prototypes of the Nanobeacon, eight in total, were installed on the Neutrino Mediterranean Observatory (NEMO) Phase II prototype tower~\cite{nemo} and three more were integrated on the prototype DU of KM3NeT~\cite{proto}. Since then, 600 Nanobeacon boards have been produced for KM3NeT, of which at present more than 200 are in operation in the deep sea. All Nanobeacons incorporate the same LED model, the HLMP-CB1A-XY0DD. The company in charge of the production also performed a calibration test, measuring the light emission intensity at different voltages. This is necessary since there is a noticeable difference in the light intensity between LEDs. The setup used by the company included a mechanical framework composed of two PVC pieces manufactured at the IFIC (Instituto de Física Corpuscular, Valencia, Spain)  workshop. The mechanical framework fixes the Nanobeacon to one of the PVC pieces while it encloses the Nanobeacon in complete darkness with the other piece. This upper piece, which has a little hole to allow the light exit from the Nanobeacon, has a space reserved to hold the head  (818-UV) of the Newport energy meter 1835-C. The test setup is shown in Fig.~\ref{fig:tna}. The light intensity  has been measured at 10 V, 16 V, 20 V, and 24 V. The measured values of the mean light intensity at each voltage are stored in the database of KM3NeT for later use. The mean values and the standard deviation of the 600 measurements are presented in Table~\ref{tab:tbpr}; the standard deviation values indicate significant intrinsic variations in intensity among individual LEDs.

 \subsection {Nanobeacon reliability}

Nanobeacons  are expected to remain operative on a regular basis for at least 20 years, with a variable frequency depending on the calibration needs. Currently, Nanobeacon operation is performed weekly in KM3NeT with a flashing frequency of 10 kHz and a typical duration of 10 minutes. 
\renewcommand{\thefootnote}{\alph{footnote}}
In order to qualify its reliability, a FIDES \footnote{FIDES means "faith" in Latin.} analysis has been applied to the Nanobeacon board as to most of the KM3NeT electronics boards~\cite{Fides}. The resulting FIT\footnote{FIT stands for \textit{Failure in Time} and indicates the number of failures per 10\textsuperscript{9} operation hours.} value is 10, which is a very low value thanks to the simplicity and robustness of the Nanobeacon electronics, corresponding to a life expectancy longer than 20 years. Moreover, the Nanobeacon will be working only during dedicated calibration runs, which represent a small fraction of the detector lifetime. In addition, since every DOM includes a Nanobeacon, a high level of redundancy is assured as a single Nanobeacon is able to illuminate several nearby DOMs.

\begin{table}[tbp]
\caption{Results of the intensity test performed after the first batch of the Nanobeacon production. The table presents the mean and the standard deviation of the intensity measurements performed on the 600 produced Nanobeacons. Note the large values of the standard deviations.}
\label{tab:tbpr}
\smallskip
\centering
\begin{tabular}{|c||c|c|c|c|}

  \hline 
 \multicolumn{1}{|c|}{ Voltage (V)} & \multicolumn {1}{|c|}{ 10}& \multicolumn {1}{|c|}{ 16 } & \multicolumn {1}{|c|}{ 20 }& \multicolumn {1}{|c|}{ 24 }\\
\hline 
  \hline 
 \multicolumn{1}{|c|}{  Mean intensity (mW) } & \multicolumn {1}{|c|}{ 2.3}& \multicolumn {1}{|c|}{ 22.2} & \multicolumn {1}{|c|}{ 49.7}& \multicolumn {1}{|c|}{85.3 }\\

 \hline 
 \multicolumn{1}{|c|}{  Standard Deviation (mW) } & \multicolumn {1}{|c|}{ 3.8}& \multicolumn {1}{|c|}{ 18.8} & \multicolumn {1}{|c|}{ 30.3}& \multicolumn {1}{|c|}{41.7}\\

 \hline
\end{tabular}

\end{table}

\begin{figure}[tbp] 
\begin{center}

\begin{subfigure}[p]{.75\linewidth}
\includegraphics[angle=0,origin=c,trim={4cm 2cm 0cm 0},clip,width=1\textwidth]{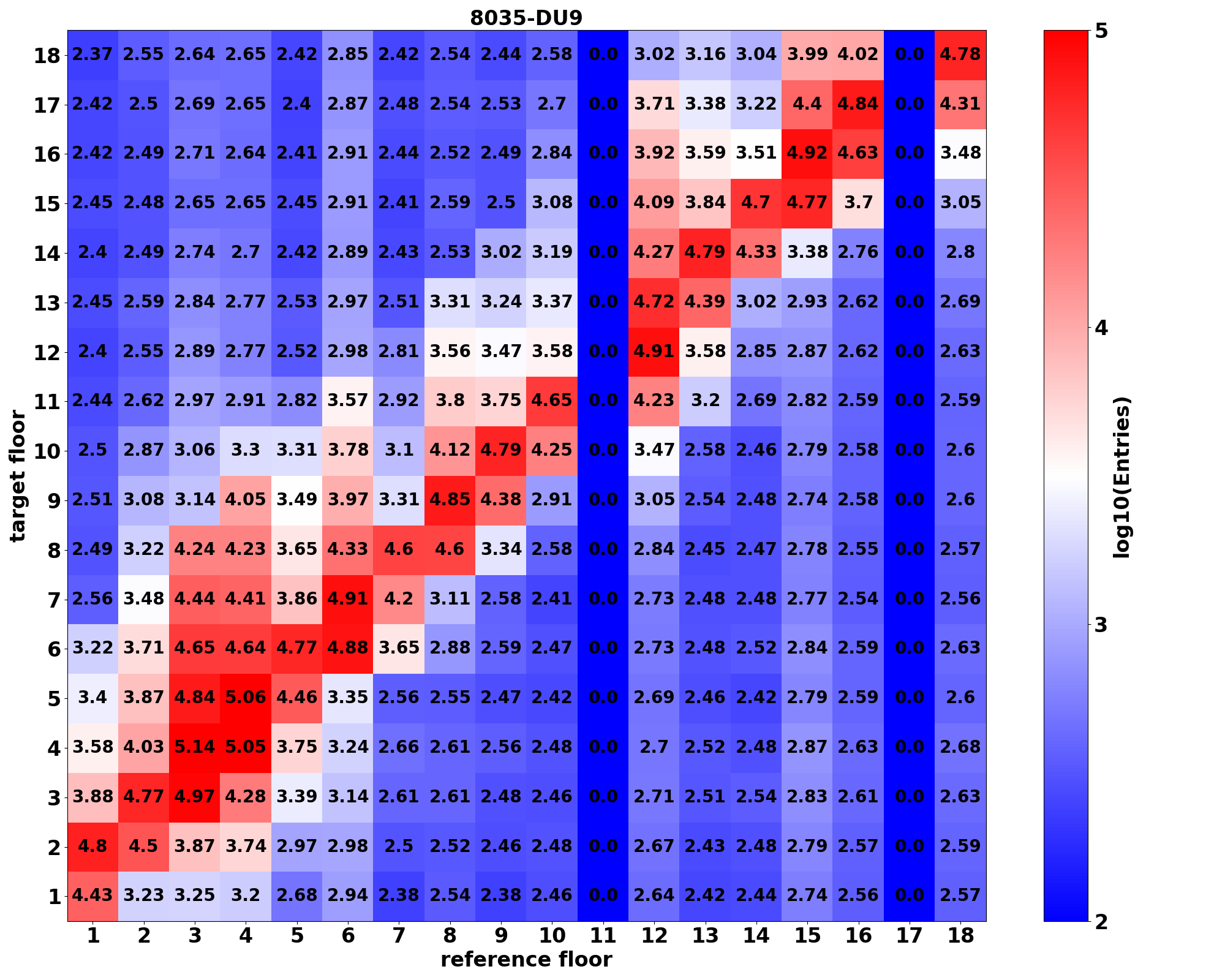}
\caption{}
\label{fig:NB_DU}
\end{subfigure}
\begin{subfigure}[p]{.75\linewidth}
\includegraphics[trim={4cm 0cm 0cm 0cm},width=1\textwidth]{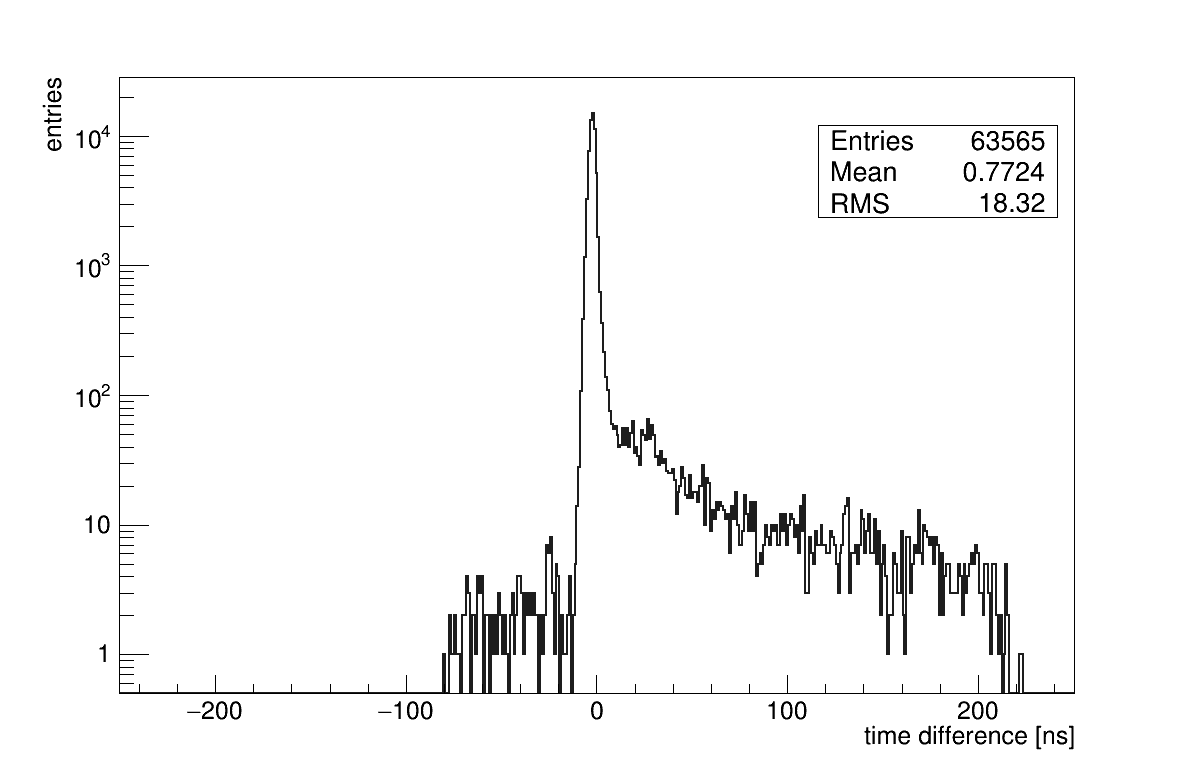}  
\caption{}
 \label{fig:domzoom}
\end{subfigure}

\caption{(a) Schematic view of the result of a Nanobeacon calibration run in KM3NeT. The x-axis shows the reference storey, i.e., the position of the light-emitting DOM in the DU (1 closest to the see storey, 18 on top); the y-axis the target storey, i.e., the position of the light-receiving DOM in the DU. The Nanobeacons were flashed sequentially with a delay of 3520~ns between one storey and the following. All Nanobeacons were set to a default voltage (8 V). The plot shows that a total coverage of the detector can be achieved with high redundancy. Therefore, the Nanobeacons have the ability of monitoring the time offsets of the PMTs in the same DOM, and in different DOMs in the same DU. For this particular run the DOMs at storey 11 and 17 were not operative. (b) Time difference between the signal received in the target DOM (second storey) and light emission in the reference DOM (first storey), corrected for the light propagation time in seawater.}
\label{fig:run}
\end{center}
\end{figure}

\begin{figure}[tbp] 

\begin{center}
\includegraphics[angle=0,origin=c,trim={0cm 5cm 0cm 
2cm},clip,width=.51\textwidth]{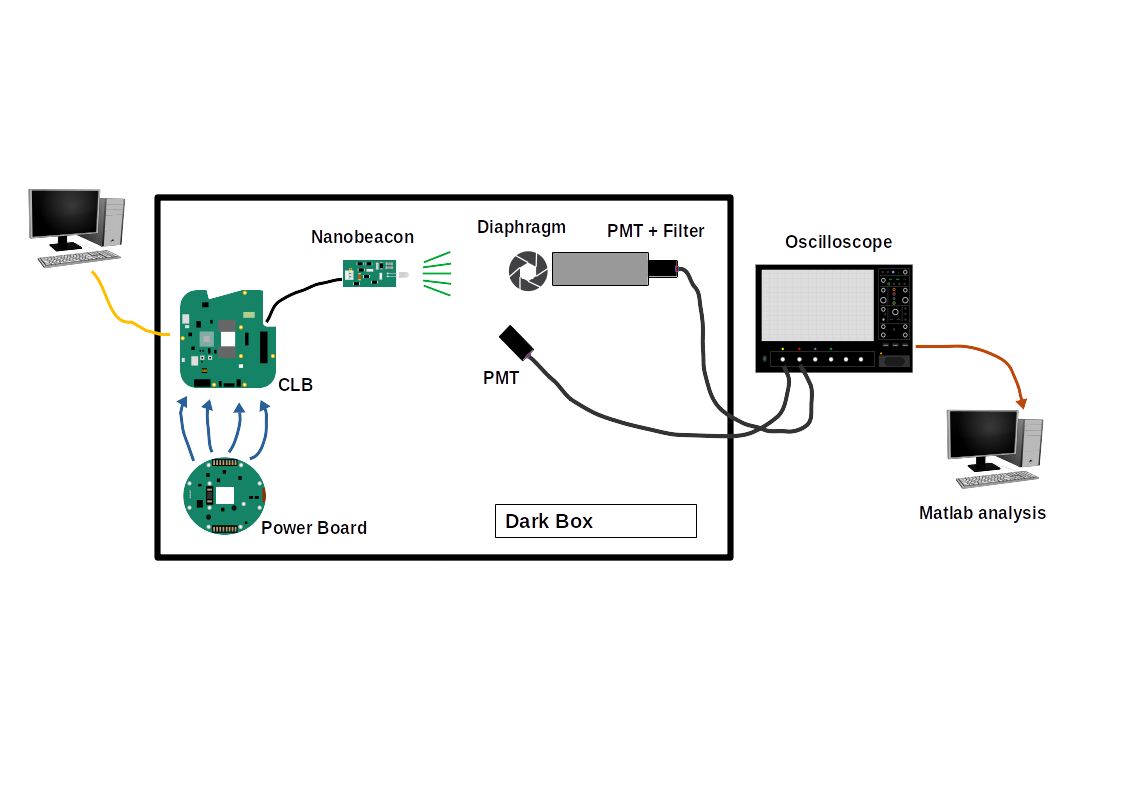}
\caption{Testbench for the measurement of the optical pulse. Inside a dark box, a Nanobeacon controlled by a CLB is flashed. Two PMTs are used. One (a Hamamatsu H653 PMT with a transit time spread lower than 200 ps)  with a filter in order to attenuate the signal to a level corresponding to one Single Photo Electron (SPE). Then, the other PMT, (a Hamamatsu H6780-03 PMT with a photo-cathode of 8 mm diameter, a rise time of 0.8 ns and a transit time of 5.4 ns) receives the pulse directly to be used as time reference. The output of both PMTs is conducted to one digital oscilloscope where the skew between both signals is measured.}


\label{fig:testsetup}
\end{center}
\end{figure}

\begin{figure}[tbp] 
\begin{center}
\includegraphics[angle=0,origin=c,trim={0.5cm 8.5cm 0.5cm 10cm},clip,width=0.49\textwidth]{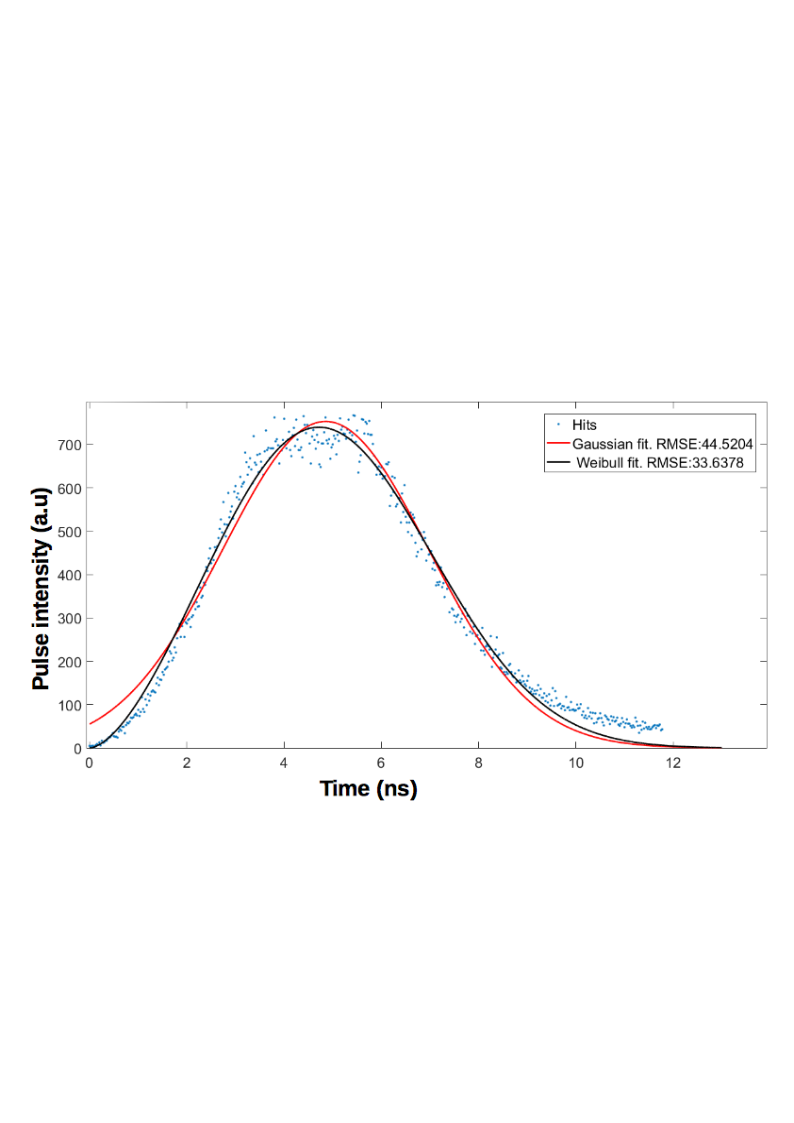}
\caption{Optical pulse shape obtained with the SPE (Single Photo-Electron) method fitted with the Gaussian and Weibull functions. Note that the Weibull function fits the experimental data  better, especially at the rise time region.}

\label{fig:nwei}
\end{center}
\end{figure}

\section{Proof of concept}\label {sec:proof}

The time calibration potential of the Nanobeacon in terms of volume coverage and redundancy is demonstrated in Fig.~\ref{fig:run}. Data from a calibration run taken in May 2020 have been used, with all the Nanobeacons in the line flashed sequentially with an operating voltage of 8 V. From the number of entries in the plot one can deduce that all Nanobeacons provide comparable intensities. The flashing frequency was 1kHz.

\section{Optical pulse characteristics}\label {sec:optical}

The rise time and width of the optical pulse shape and the spectrum of the emitted light must be characterized in order to evaluate their effect on the time calibration. In this section, the results of the qualification  are presented, together with a brief description of the methods used.  


\subsection {Pulse shape}
\subsubsection {Single photoelectron method}
The SPE technique~\cite{photo} has been chosen to measure the rise time and the width of the Nanobeacon emitted pulse. Fig.~\ref{fig:testsetup} shows the scheme of the measurement setup. The technique allows for the reconstruction of the pulse shape independently of the gain and rise time of the detector. A light sensor, usually a PMT with a low Transit Time Spread (TTS) and enough sensitivity to detect single photons, is needed. In this case, two detectors have been used in the experimental setup: a Hamamatsu H6780-03 PMT with a photo-cathode of 8 mm diameter, a rise time of 0.8 ns and a transit time of 5.4 ns is used as trigger and provides an electrical signal generated by the first photons arriving from the optical pulse. A second detector,  a Hamamatsu H653 PMT with a transit time spread lower than 200 ps and high sensitivity, measures the arrival time of the photons in the attenuated pulse. The Nanobeacon pulse is attenuated by means of optical filters so that only one or no photoelectron per pulse reaches the PMT. To ensure the PMT only detects single photon hits, a level of filtering that triggers no more than one over 100 Nanobeacon pulses is set. In this way, the probability to have no signal at all is 99\%. Assuming Poissonian statistics, it is straightforward to estimate that the average number of single photoelectrons per flash is lower than $\mu$ = 0.01, and the probability to have two single photoelectrons in the same flash is negligible (0.005\%). 


The signals of both PMTs are measured by a digital oscilloscope with a high bandwidth and sampling period. The time differences between the trigger and the SPE PMT are measured and accumulated in a histogram that reproduces the optical pulse shape and allows the measurement of the rise time and the FWHM of the pulse.  The distribution of the time differences reproduces the optical shape because the probability that a single photon arrives at the SPE PMT is related to the shape of the optical pulse, i.e., photons in the center of the optical pulse have higher probability to be detected than those in the leading or trailing region of the optical pulse. This technique has numerous advantages with respect to the use of PMTs of high precision. First of all, the SPE technique is based on the measurement of the temporal differences of two signals, greatly reducing the residual errors. Moreover, the digitization of the signals is very effective and has higher time resolution if a Digital Oscilloscope with high bandwidth, sampling period and large memory is used. Large memory is specially useful in order to acquire the sufficient number of hits and increase the precision of the measurement. Finally, high precision PMTs are considerably more expensive than the PMTs needed to perform the SPE technique.

\subsubsection {Results}

In order to measure the rise time and the width of the LED model chosen for KM3NeT, the pulse shape of 24 LEDs was measured using the SPE technique. The pulse shape was measured for each LED using a voltage of 24 V and a frequency of 1 kHz. The time distribution obtained for one of the LEDs using the SPE technique is shown in Fig.~\ref{fig:nwei}. The results of the fits to Gaussian and Weibull functions are superimposed. In Table~\ref{tab:tbp} the mean and the standard deviation of the rise time and the width (FWHM) of the 24 LEDs using a Gaussian and a Weibull function is shown. The rise time and the FWHM of the optical pulses are about 2.69 and 4.51 ns respectively, with a statistical accuracy around 8-9\%. From the figure, it can be appreciated that the Weibull function matches the data better, in particular at the beginning of the pulse, providing a more precise description of the rise time of the pulse.  

\begin{figure*}[tbp] 

\begin{center}
\includegraphics[angle=0,origin=c,trim={4cm 0cm 0cm 0cm},clip,width=1.1\textwidth]{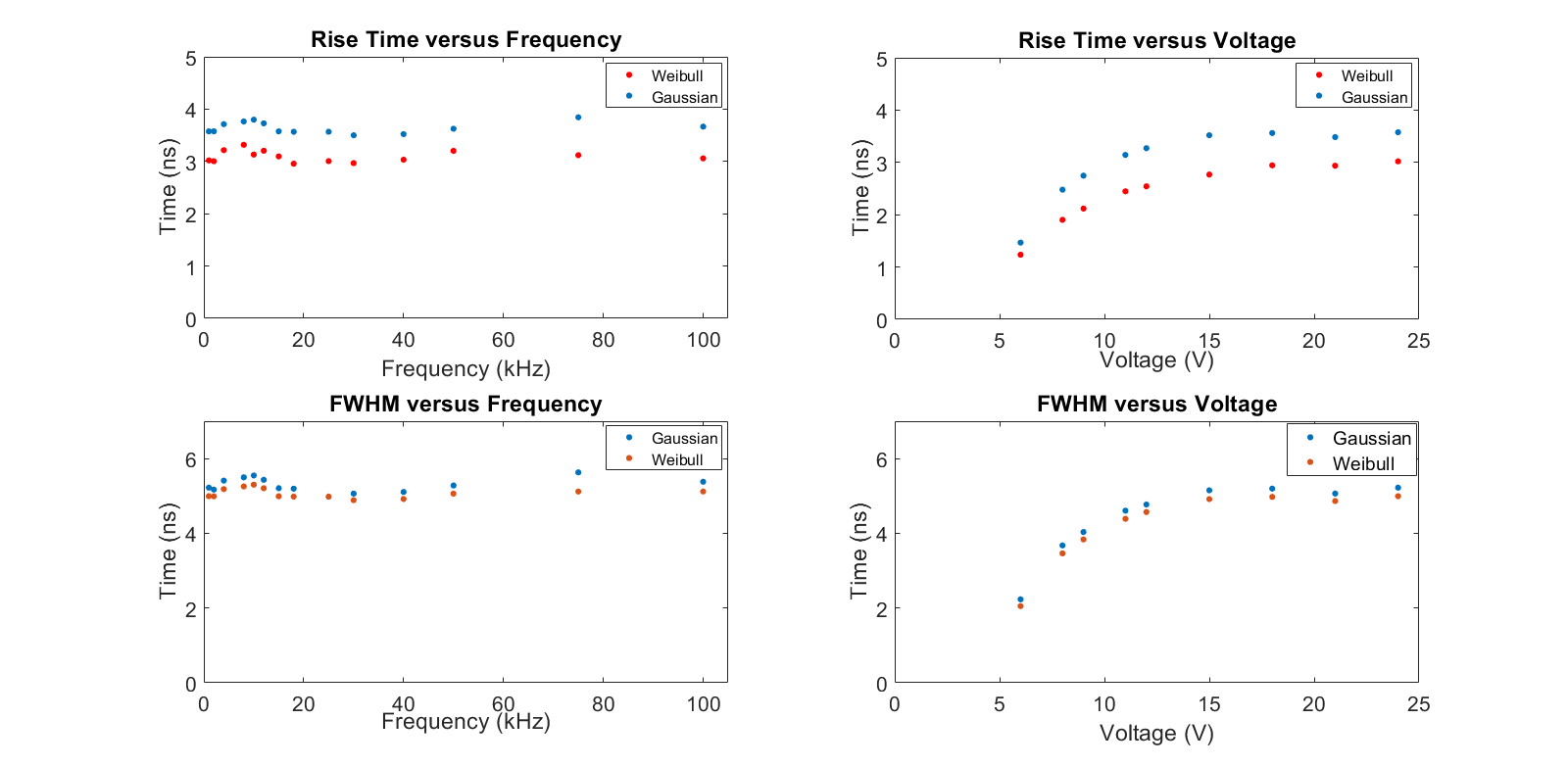}
\caption{ Rise time and width (FWHM), obtained from Gaussian and Weibull
fits, as a function of flashing frequency and applied voltage. When varying the frequency the voltage is fixed at 24 V. When varying the voltage the frequency is fixed at 1 kHz. Note the correlation of the voltage with the pulse width.}

\label{fig:vff}
\end{center}
\end{figure*}

\begin{table}[tbp]
\caption{Mean values of the rise time and the FWHM for a Gaussian and a Weibull fit. The results for 24 LEDs of the model HLMP-CB1A-XY0DD being powered at 24 V and triggered at 1 kHz are presented.}
\label{tab:tbp}
\smallskip
\centering
\begin{tabular}{ |p{2cm}|p{1cm}|p{1cm}|p{0.2cm}|p{0.1cm}|}

  \hline 
 \multicolumn{1}{|c|}{  N LEDs } & \multicolumn {2}{|c|}{ Rise time (ns)}& \multicolumn {2}{|c|}{ FWHM (ns)} \\

 & \multicolumn {1}{p{1cm}}{ Gaussian}& \multicolumn {1}{p{0.1cm}}{Weibull} \vline& \multicolumn {1}{p{1cm}}{ Gaussian}& \multicolumn {1}{p{1cm}}{Weibull} \vline \\
\hline

\centering Mean value &\centering 3.14&\centering 2.69& \centering 4.39 &\multicolumn{1}{|c|}{4.51}\\
\hline
\centering Standard deviation &\centering 0.27 (8.6\%) &\centering 0.21 (7.8\%)& \centering 0.38 (8.6\%) &\multicolumn{1}{p{1.1cm}}{0.41 (9.1\%)}\vline \\
 \hline

 \hline
\end{tabular}
\end{table}

\begin{figure}[tbp] 
\begin{center}

\includegraphics[angle=0,origin=c,trim={1.5cm 9.5cm -0.5cm 10cm},clip,width=0.54\textwidth]{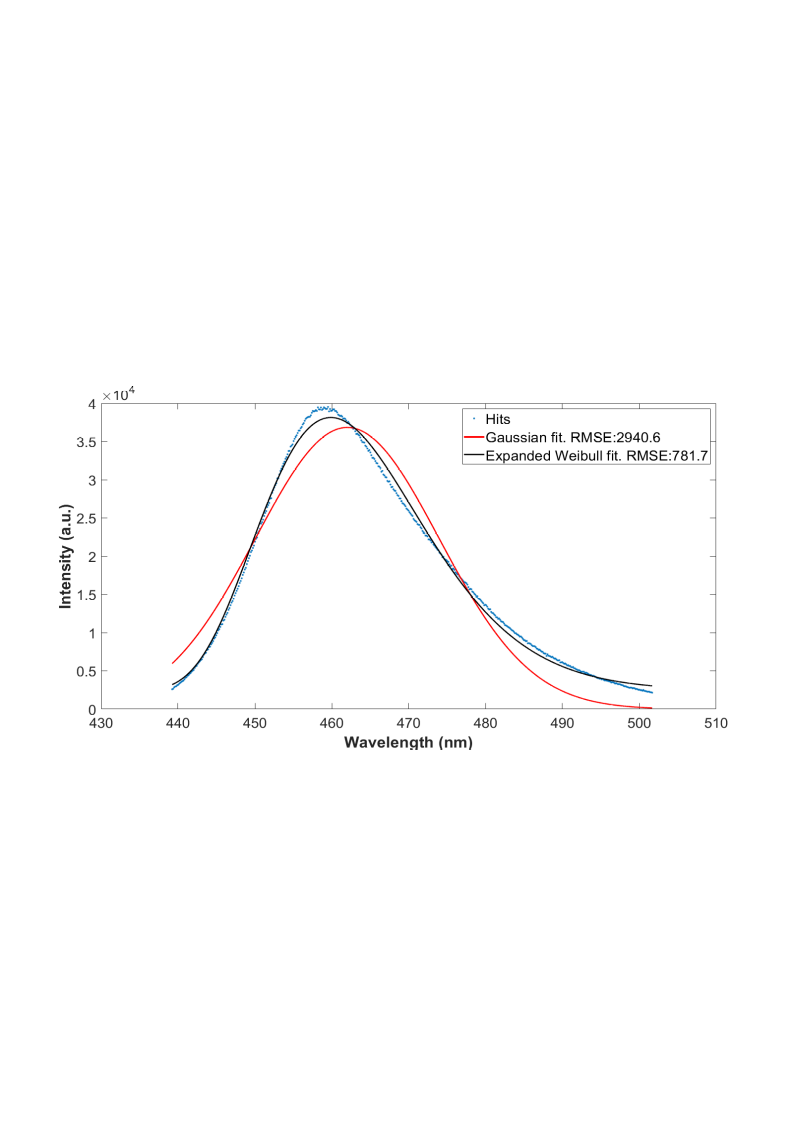}
\caption{Wavelength spectrum of an LED operated at 30V and 1 kHz. Both, a Gaussian and a Weibull fit are included.}

\label{fig:nwl2}
\end{center}
\end{figure}

\begin{figure}[tbp] 
\begin{center}

\includegraphics[angle=0,origin=c,trim={1.5cm 9.5cm -0.5cm 10cm},clip,width=0.54\textwidth]{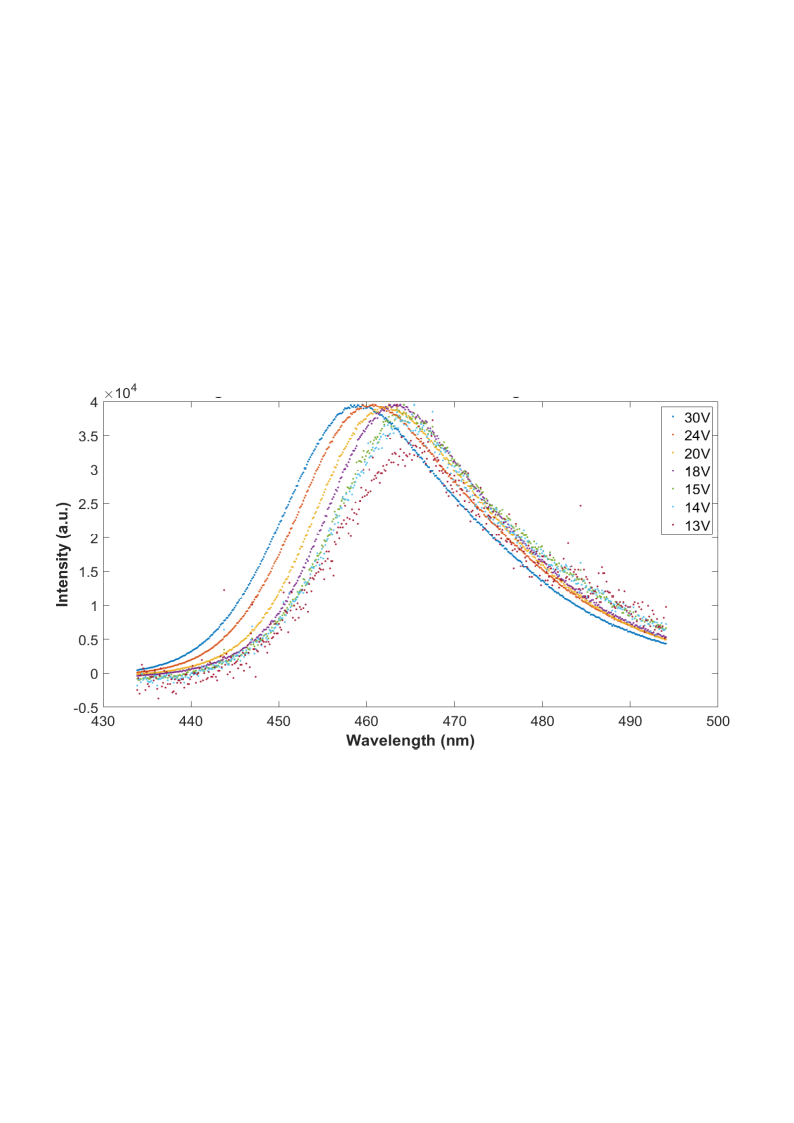}
\caption{Wavelength spectra for different voltage values. The data are normalized  to the maximum intensity. The wavelength distribution shifts to higher wavelengths when the voltage is reduced.}

\label{fig:lwl}
\end{center}
\end{figure}

\begin{figure}[tbp] 
\begin{center}

\includegraphics[angle=0,origin=c,trim={0cm 10cm 0cm 10cm},clip,width=0.54\textwidth]{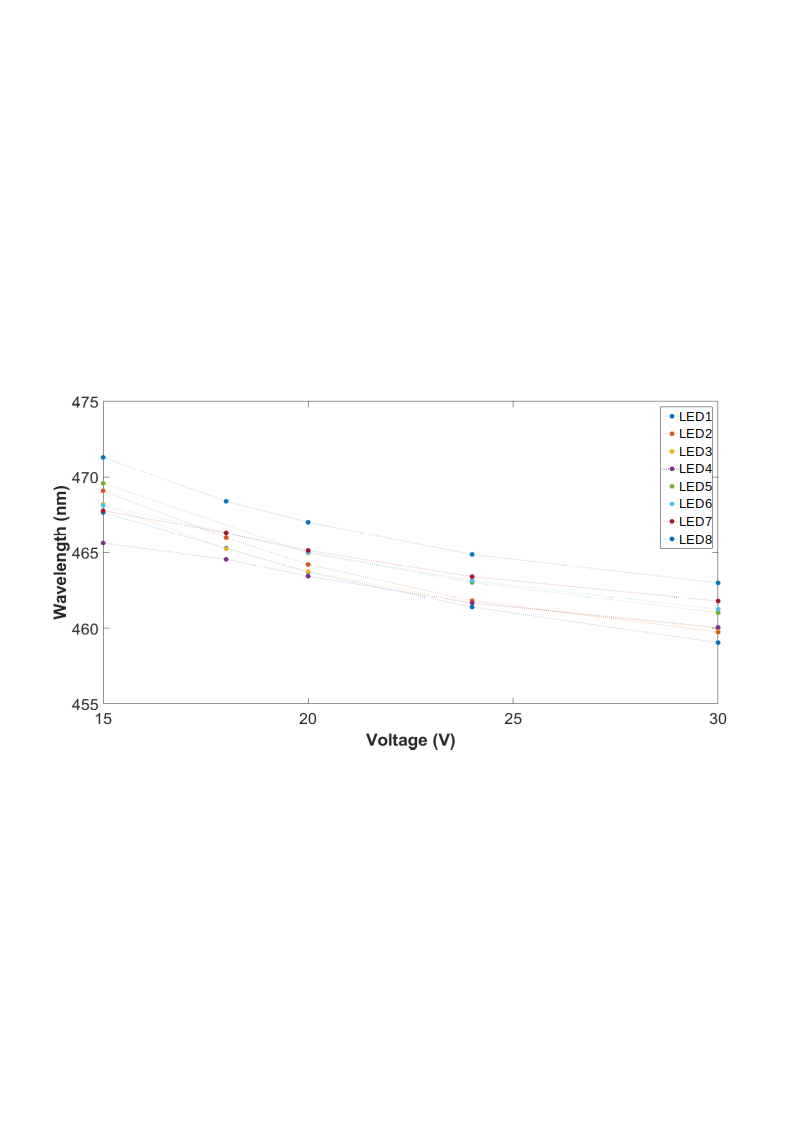}
\caption{Change of the central wavelength, calculated with the Weibull fit, as a function of the voltage applied  for a sample of 8 LEDs. The central wavelengths vary slightly with the voltage, increasing as the voltage decreases.}

\label{fig:nwl}
\end{center}
\end{figure}

\subsubsection {Effect of voltage and flashing frequency on the pulse shape}

Rise time and FWHM of the optical pulse have been measured at different voltages and frequencies in order to evaluate the impact of these two parameters in the shape of the optical pulse. Each set of LED data has been fitted using a Weibull and a Gaussian function. The voltages have been varied between 6 V and 24 V, while maintaining a fixed
frequency of 1 kHz. In another set of tests, the frequency has been varied from 1 kHz up to 100 kHz keeping the voltage fixed at 24 V. All the tests have been carried out using the same test equipment (Electronics + Nanobeacon + LED). As can be seen in Fig.~\ref{fig:vff}, the impact of the frequency on the rise time and the FWHM of the optical pulse is negligible, and the fluctuations are compatible with the measurement uncertainties. However, the voltage has a clear impact on both rise time and FWHM. At low voltages (6 V), both decrease to levels only half of the values measured at higher voltages. The lower the voltage, the fainter and narrower the optical pulse, giving lower values of rise time and FWHM. This is especially true for voltages below 15 V. Note that the minimum rise time of 1.24 ns (Weibull fit) at 6 V is very close to the rise time of the electrical trigger signal. The mean of the rise time of the electrical trigger signal, after 1100 electrical pulses have been measured, has been computed to be 1.34 ns with a standard deviation of 50 ps.

\subsection{Wavelength characteristics} \label {sec:wavelength}

The wavelength spectrum of the optical pulse is of great importance as the water optical properties depend on the wavelength, therefore it has an impact on the travel time of the pulse and in the time calibration.  The wavelength spectrum of the optical pulse was measured for a set of 8 LEDs with an imaging spectrometer (Triax 190 de Horiba Scientific) at different voltages. The LEDs were tested at 30 V, 24 V, 20 V, 18 V, 15 V, 14 V and 13 V, keeping the frequency fixed at 1 kHz. Changes in the frequency do not affect the wavelength distribution. The spectral wavelength of an LED is shown in Fig.~\ref{fig:nwl2}, where a Gaussian and a Weibull fit are superimposed. The wavelength spectra for different voltage values are shown in Fig.~\ref{fig:lwl}. The wavelength spectrum distribution shape does not change, but the wavelength peak slightly moves towards lower values when the voltage increases. This can also be seen graphically in Fig.~\ref{fig:nwl}.  On average, the distribution shifts 2 nm to higher wavelength values when the voltage decreases by 5 V, this change
being larger for lower voltages. A variation of 10 nm in the light wavelength corresponds to a change in the refraction index lower than 0.06 \%~\cite{refra}. This allows for the use of the Nanobeacon as time accurate (ns resolution) source over pathlengths higher than 100 m. 



\section{Summary} \label {sec:conclu}

In order to ensure the high angular resolution required for the reconstruction of the neutrino direction in KM3NeT, a relative time synchronization of the order of 1 ns between DOMs is needed. Different time calibration systems have been developed to this end.  One of them is the Nanobeacon device which is presented in this article. The time properties of the basic electronics together with the firmware used for the operation of the board are described. The properties  (rise time and width) of the optical pulse generated by the Nanobeacon have been evaluated, taking into account the impact of voltage and frequency on its performance. The wavelength spectrum of the optical pulses has also been measured,  as well as the effect of the input voltage. The results obtained show that the Nanobeacon is an excellent instrument, with low cost while fulfilling the time calibration requirements of KM3NeT. The characterization performed, measuring the optical pulse and its wavelength distribution, will be of great help during KM3NeT operation. The future plans include the analysis of LED models with different wavelength (ranging from near ultraviolet to green) and higher intensity that could be deployed in KM3NeT for water properties characterization.

\section*{Acknowledgement} 
The authors acknowledge the financial support of the funding agencies: Agence Nationale de la Recherche (contract ANR-15-CE31-0020), Centre National de la Recherche Scientifique (CNRS), Commission Europ\'eenne (FEDER fund and Marie Curie Program), Institut Universitaire de France (IUF), LabEx UnivEarthS (ANR-10-LABX-0023 and ANR-18-IDEX-0001), Paris \^Ile-de-France Region, France; Shota Rustaveli National Science Foundation of Georgia (SRNSFG, FR-18-1268), Georgia; Deutsche Forschungsgemeinschaft (DFG), Germany; The General Secretariat of Research and Technology (GSRT), Greece; Istituto Nazionale di Fisica Nucleare (INFN), Ministero dell'Universit\`a e della Ricerca (MIUR), PRIN 2017 program (Grant NAT-NET 2017W4HA7S) Italy; Ministry of Higher Education Scientific Research and Professional Training, ICTP through Grant AF-13, Morocco; Nederlandse organisatie voor Wetenschappelijk Onderzoek (NWO), the Netherlands; The National Science Centre, Poland (2015/18/E/ST2/00758); National Authority for Scientific Research (ANCS), Romania; Ministerio de Ciencia, Innovaci\'{o}n, Investigaci\'{o}n y Universidades (MCIU): Programa Estatal de Generaci\'{o}n de Conocimiento (refs. PGC2018-096663-B-C41, -A-C42, -B-C43, -B-C44) (MCIU/FEDER), Generalitat Valenciana: Prometeo (PROMETEO/2020/019), Grisol\'{i}a (ref. GRISOLIA/2018/119) and GenT (refs. CIDEGENT/2018/034, /2019/043, /2020/049) programs, Junta de Andaluc\'{i}a (ref. A-FQM-053-UGR18), La Caixa Foundation (ref. LCF/BQ/IN17/11620019), EU: MSC program (ref. 101025085), Spain.


%





\ifCLASSOPTIONcaptionsoff
  \newpage
\fi





\bibliographystyle{IEEEtran}

%

\begin{IEEEbiography}{Diego Real}
is a PhD. in Physics and Research Engineer at Instituto de F\'isica Corpuscular. He received his BS in Electronics in 1997 and his MS in Control and Electronics in 2000, both from the Polytechnic University of Valencia.  He is the author of several publications on electronics. His current research interests include acquisition and synchronization systems for particle physics. He is, since 2013, the Electronics project leader of the KM3NeT telescope and member of the Technical Advisory Board of the GVD-Baikal telescope.
\end{IEEEbiography}
\begin{IEEEbiography}{David Calvo} 
is a PhD. in Physics and research engineer at Instituto de F\'isica Corpuscular of Valencia. He received his MS in Computing in 2006 from University Jaume I, his MS in Electronics in 2009 from University of Valencia and his MS in electronic systems design in 2012 from the Polytechnic University of Valencia. His research interests are focused on the digital electronics, synchronization and readout acquisition systems. He is the author of several publications on electronics.
\end{IEEEbiography}

\begin{IEEEbiography}{Francisco Salesa}
is a Distinguished Researcher at Institut de Física Corpuscular (IFIC) since 2019. He received his Physics Degree from University of Valencia in 2003, his master degree in 2006, and his PhD in Physics in 2010 with the title “Time Calibration and Point Source Analysis with the ANTARES Neutrino Telescope”, work which was awarded in 2012 with the “Premi Extraordinari de Doctorat” for outstanding theses. He had several post-doctoral positions at Colorado State University (US) 2011-2013, Penn State University (US) 2013-2015, and Institute of Nuclear Physics (Poland) 2015-2019. He has worked for The Pierre Auger and The HAWC Gamma-ray Observatories, where he was in charge of the calibration of the detector. He is the author of several publications on detector calibrations. His research interests include multi-messenger astronomy and the search for sources of high-energy cosmic rays. 

\end{IEEEbiography}





\vfill


\end{document}